\documentstyle[preprint,floats,graphics,tighten,aps,eqsecnum]{revtex}              
\begin{document}


\def\lqcd{\Lambda_{\rm QCD}}
\def\xslash#1{{\rlap{$#1$}/}}
\def\dsl{\,\raise.15ex\hbox{/}\mkern-13.5mu D}
\preprint{\vbox{\hbox{UTPT--01-02}
\hbox{hep-ph/0101190}}}

\def\ctp#1#2#3{\CTP{\bf #1} (#2) #3}
\def\jetpl#1#2#3{\JETPL{\bf #1} (#2) #3}
\def\nc#1#2#3{\NC{\bf #1} (#2) #3}
\def\np#1#2#3{\NP{\bf B#1} (#2) #3}
\def\pl#1#2#3{\PL B {\bf #1} (#2) #3}
\def\prl#1#2#3{\PRL{\bf #1} (#2) #3}
\def\prd#1#2#3{\PR D {\bf #1} (#2) #3}
\def\prep#1#2#3{\PRep{\bf #1} (#2) #3}
\def\physrev#1#2#3{\PR{\bf #1} (#2) #3}
\def\sjnp#1#2#3{\SJNP{\bf #1} (#2) #3}
\def\nuvc#1#2#3{\NC{\bf #1A} (#2) #3}
\def\blankref#1#2#3{   {\bf #1} (#2) #3}
\def\ibid#1#2#3{{\it ibid,\/}  {\bf #1} (#2) #3}
\def\AP{{\it Ann.\ Phys.\ }}
\def\CMP{{\it Comm.\ Math.\ Phys.\ }}
\def\CTP{{\it Comm.\ Theor.\ Phys.\ }}
\def\IJMP{{\it Int.\ Jour.\ Mod.\ Phys.\ }}
\def\JETPL{{JETP Lett.\ }}
\def\NC{{\it Nuovo Cimento\ }}
\def\NP{{Nucl.\ Phys.\ }}
\def\PL{{Phys.\ Lett.\ }}
\def\PR{{Phys.\ Rev.\ }}
\def\PRep{{Phys.\ Rep.\ }}
\def\PRL{{Phys.\ Rev.\ Lett.\ }}

\title{Renormalon Analysis of Heavy-Light Exclusive $B$ Decays}

\author{Craig N. Burrell and Alexander R. Williamson}

\address{ \medskip
	Department of Physics, University of Toronto\\
	60 St.~George Street, Toronto, Ontario,
	Canada M5S 1A7 \medskip}


\maketitle

\begin{abstract}
We study two-body exclusive decays of the form 
$\bar{B} \to D^{(*)} L \; (L = \pi, \rho)$ in the heavy-quark limit.
We perform a renormalon analysis of such processes to determine
the order at which nonperturbative factorization-breaking power corrections
enter the amplitude. 
We find that a class of leading power 
corrections to the color octet matrix element, of 
${\cal O}(\lqcd/m_b)$, vanish in the limit of a symmetric light meson parton distribution
function.  We
discuss the phenomenological significance of this result.
\end{abstract}

\pagebreak

\section{Introduction}
The weak decays of $B$ mesons into hadronic final states are important for 
an understanding of the CKM sector of the standard model, and particularly
for the study of CP violation.  These decays involve a mixture of
calculable weak physics, perturbative QCD, and nonperturbative QCD.  It is 
the latter component which contains the bulk of the theoretical difficulty, and it
is primarily contained in the evaluation of low energy matrix elements of quark operators.  

A common assumption used to simplify the nonperturbative
component of hadronic decays is 
factorization, which specifies that the hadronic matrix element
of a four-quark operator be factored into two matrix elements of
simple currents.  For example,
\begin{equation}
\label{naive_factorization}
\langle \pi^- D^+ | (\bar{c}b)_{V-A}(\bar{d}u)_{V-A}|\bar{B}\rangle \to
 \langle \pi^- |(\bar{d}u)_{V-A}|0 \rangle \langle D^+|(\bar{c}b)_{V-A}|\bar{B}\rangle.
\end{equation}
This prescription (which, following the authors of \cite{bbns1,bbns2}, we refer to as `naive factorization') 
considerably simplifies matters because the factored structures
on the right hand side may be parameterized in terms of decay constants and form
factors.  It amounts, however, to ignoring corrections which 
connect the $(\pi^-)$ to the $(\bar{B} D^+)$ system.  Since these corrections are
responsible for final-state rescattering and strong interaction phase shifts,
leaving them out ignores important physics.  Also, the left hand side of 
(\ref{naive_factorization}) is
renormalization scale dependent, while the right-hand side is not --- a clear
indication that relevant physics is being lost.  

Recently it was argued that, for certain $B$ decays to heavy-light final states
$\bar{B} \to H L$, the strong interactions which break factorization are 
hard in the heavy quark
limit \cite{bbns1,bbns2}, and can therefore be calculated perturbatively.  This
proposal has been explicitly verified to two-loop order \cite{bbns2}.  
This idea allows one to include perturbative corrections missing in 
(\ref{naive_factorization})
without introducing any new nonperturbative parameters.
It thus provides a remarkably attractive means to
study rescattering and strong phases in two-body hadronic decays. A generalization
of this idea has also been proposed for decays to two light mesons
\cite{bbns1,bbns2}, but
in this article we will restrict our attention to final states with one heavy 
meson $(D,D^*)$ and one light meson $(\pi,\rho)$.

A potential problem with this proposal is that it is valid only in 
the strict heavy quark limit \footnote{Following the authors of \cite{bbns1,bbns2},
we assume that the physical $b$ quark mass is not so large that Sudakov form factors
modify the power counting.}.  It receives power corrections of the form 
$(\lqcd/m_b)^n$ from a variety of sources: examples are 
hard spectator interactions, non-factorizable soft and collinear gluon
exchange, and transverse momenta of quarks in the light meson.
Unlike power corrections in inclusive $B$ decays, there is as yet no
systematic way to compute these corrections for exclusive decays.  By naive power
counting one expects such corrections at ${\cal O}(\lqcd/m_b)$, but situations are 
known where the naively expected corrections vanish. For instance, 
in the zero recoil $B \to D$
transition matrix element in Heavy Quark Effective Theory the leading
$1/m_b$ corrections vanish \cite{luke}.
Therefore rather than relying on the naive expectation, one would like
to calculate the power corrections directly.

In the absence of a general theory of power corrections, we aim to 
determine at least the order at which power corrections enter.  
This may be done by carrying out a renormalon analysis,
which involves calculating a subset of Feynman graphs at each order in
perturbation theory. It exploits the fact that the perturbative series 
in quantum field theory is asymptotic, and permits one to extract from the large-order
behavior of the theory information about the scaling behavior of
nonperturbative power corrections.  
In this paper, we use renormalons to assess the parametric size of a subset of
power corrections to the separation (\ref{naive_factorization}) for the class of
$\bar{B} \to H L$ decays discussed in \cite{bbns2}.  

The remainder of this paper is organized as follows: in Section \ref{renormalonTheory} we review the 
theoretical framework of renormalon analyses.  In Section \ref{BDecays} we discuss the phenomenological 
context and motivation of our study.
In Section \ref{calculation} we describe the renormalon 
calculation and give the result.  Section
\ref{conclusions} contains our conclusions. 

\section{Renormalons}
\label{renormalonTheory}

A typical amplitude in QCD perturbation theory may be expressed in the general form
\begin{equation}
\label{Rseries}
R(\alpha_s) = \sum_{n=0}^{\infty} R_n \alpha_s^{n+1}.
\end{equation}
Normally one only
calculates a few terms in this series.  However, one may ask about the general behavior of
this series at large orders in perturbation theory.  It has been argued 
\cite{beneke99} that quantum field
theories of phenomenological interest have large order coefficients of the form
\begin{equation}
R_n \stackrel{n\to\infty}{\sim} a^n n! \, n^b
\end{equation}
for some constants $a$ and $b$.  Clearly, such a series is factorially divergent.
It might appear that, as a result, a sum for the series cannot be defined.  However, one
may define the Borel sum $\tilde{R}$ in the following way. Perform a Borel transformation on the
series:
\begin{equation}
\label{Btransform}
R(\alpha_s) = \sum_{n=0}^{\infty} R_n \alpha_s^{n+1}  \Longrightarrow B[R](t) = 
\sum_{n=0}^{\infty} \frac{R_n}{n!} t^n .
\end{equation}
This series in terms of the Borel parameter $t$ is convergent and may be explicitly summed.  
One may then perform an inverse Borel transformation to obtain the Borel sum
\begin{equation}
\label{inverseBtransform}
\tilde{R} = \int_0^\infty \; dt \; e^{-t/\alpha_s} B[R](t).
\end{equation}
The original series $R$ and the Borel sum $\tilde{R}$ have the same series expansion.  

In some cases, the transformed series $B[R](t)$ has poles along the positive real axis 
\cite{beneke99}.  When these
poles are encountered in the inverse Borel transformation (\ref{inverseBtransform}), one is forced to
deform the integration contour either above or below the real axis. Nothing specifies which choice to make, 
yet the result of the integration 
depends on the choice. As a result, the Borel sum acquires an ambiguity.
For a simple pole located at $t_0 > 0$, the ambiguity is 
\begin{eqnarray}
\delta \tilde{R} &\sim& e^{-t_0/\alpha_s(\mu)} \nonumber \\
 &\sim& \left( \frac{\lqcd}{\mu} \right)^{-2 \beta_0 t_0} = \left( \frac{\lqcd}{\mu} \right)^{2u_0}
\label{ambiguity}
\end{eqnarray}
where we have defined $u_0 = - \beta_0 t_0 > 0$.
The ambiguity has the form of a nonperturbative power correction.  For physical quantities, which
cannot be ambiguous, there must be present power corrections to remove this ambiguity.  Note that for
$\mu > \lqcd$ it is the pole nearest the origin which gives the leading power correction.  
This simple sketch illustrates how the large-order perturbative behavior of 
the theory reveals something about the nonperturbative sector of the theory \cite{beneke99}.

In general, this approach only permits one to determine the order of the power 
corrections and not their coefficients or analytic form.  
Furthermore, while a pole at $u_0$ in the Borel plane definitely indicates the presence of 
power corrections $\sim (\lqcd/m_b)^{2 u_0}$, the absence of a pole does not 
necessarily imply the absence of power corrections of that order.  The absence of a pole
is, rather, suggestive that power corrections of the corresponding order are 
absent \cite{beneke99}.
The renormalon technique has been applied 
in a variety of contexts where a general theory of power corrections has not been
available 
\cite{mueller85,manoharwise95,beneke/braun94,bsuv94,webber94,luke/manohar/savage95}.  
In Section \ref{calculation} we study power corrections to factorization in this way.

\begin{figure}[htbp]
\centerline{\scalebox{0.8}{\includegraphics{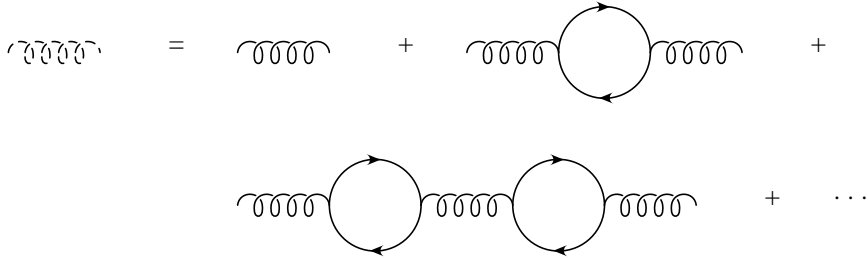}}}
\caption{The renormalon `bubble chain' propagator.} 
\label{Rprop} 
\end{figure}

 In practice one cannot sum the entire perturbative series (\ref{Rseries}) to obtain
an exact expression for the Borel transformed amplitude (\ref{Btransform}).  Instead,
one sums a subset of the Feynman diagrams at each order, implicitly taking the 
resulting analytic structure to be characteristic of the full result.  Typically,
all-orders contributions are obtained by inserting into graphs `bubble chain' 
propagators of the kind shown in 
Figure 1.  One may take the 
formal limit $N_f \to \infty$ with $\alpha_s N_f$ fixed, in which case the set of graphs
with a single `bubble chain' insertion are dominant \cite{beneke99}.
This `bubble chain' propagator, which we denote by a dashed gluon line, has the form
\cite{beneke/braun94,beneke93}
\begin{equation}
D_{\mu\nu}(k) = \frac{i}{k^2} \left( -g_{\mu\nu} + \frac{k_\mu k_\nu}{k^2} \right) \sum_{n=0}^\infty ( \beta_0 N_f \alpha_s )^n
	( \ln(-k^2/\mu^2) + C)^n.
\end{equation}
The factors in the sum arise from the fermion loops depicted in Figure \ref{Rprop}.
These loops have been renormalized
in an MS-like scheme, and $C$ is a scheme dependent constant.  In the $\overline{\rm MS}$ scheme, $C = -5/3$.  We 
discuss renormalization of amplitudes containing 
the renormalon propagator in more detail in Section \ref{calculation}.  The Borel transform of this 
propagator with respect to $\alpha_s N_f$ is
\begin{eqnarray}
\label{renProp}
B[D_{\mu\nu}(k)](u) &=& \frac{1}{\alpha_s N_f} \frac{i}{k^2} \left( -g_{\mu\nu} + \frac{k_\mu k_\nu}{k^2} \right) 
	\sum_{n=0}^\infty \frac{(-u)^n}{n!} ( \ln(-k^2/\mu^2) + C)^n \nonumber \\
	&=& \frac{1}{\alpha_s N_f} \left( \frac{\mu^2}{e^C} \right)^u 
	\frac{i}{(-k^2)^{2+u} } ( k_\mu k_\nu - k^2 g_{\mu\nu} ).
\end{eqnarray}
The limit $u \to 0$ of this expression, 
equivalent to retaining only the first term in the expansion depicted in Figure 1, reduces to the 
usual gluon propagator as expected.  For Feynman graphs in which the $\alpha_s$ dependence arises only from gluon exchange, the Borel transformed graph
is obtained by replacing the gluon propagator by this renormalon propagator.  

\section{Factorization and Two-Body Hadronic $B$ Decays}
\label{BDecays}

We work in an effective theory where the weak bosons and top quark have been 
integrated out.  The relevant part of the effective Hamiltonian, valid below $M_W$, is
\begin{equation}
\label{Heff}
{\cal H}_{eff} = \frac{G_F}{\sqrt{2}} V_{ud}^* V_{cb} (C_1 {\cal O}_1 + 
	C_8 {\cal O}_8) 
\end{equation}
where the singlet and octet operators are, respectively,
\begin{eqnarray}
\label{operators}
{\cal O}_1 &=& \bar{c} \gamma_\mu (1-\gamma_5) b\; \bar{d} \gamma^\mu (1 - \gamma_5) u ,\\
{\cal O}_8 &=& \bar{c} \gamma_\mu (1-\gamma_5) T^a b\; \bar{d} \gamma^\mu (1 - \gamma_5) T^a u.   
\end{eqnarray}
The running of the Wilson coefficients $C_i$ has been calculated at next-to-leading order \cite{italians81,burasweisz90}.  

The computation of any Feynman amplitude in this theory involves evaluating matrix elements of the 
operators ${\cal O}_{1,8}$.  In general such matrix elements contain both hard and soft physics.  One would
like to disentangle these energy scales; the hard physics could be calculated directly and the soft physics
parameterized by form factors and decay constants.  Though it is not obvious that the physics can be disentangled
in this way, it has been argued \cite{bbns1,bbns2} that in certain situations it is possible to do so.  

More
specifically, consider a decay of the form $\bar{B} \to H L$, where $H$ is a 
heavy meson $(H = D,D^*)$ and $L$ is a light meson $(L = \pi,\rho)$.  We require 
that the topology of the decay be such that the light quark in the 
initial state is transferred to the heavy final
state meson.  In this case, and in the heavy quark limit, it is argued that 
`non-factorizable' corrections are perturbative and may be calculated.  These statements
are summarized in the factorization equation \cite{bbns1,bbns2}
\begin{equation}
\label{masterequation2}
\langle H L | {\cal O}_i | \bar{B} \rangle = F^{\bar{B} \to H}(m_L^2) f_L \int_0^1 dx \,T^I_i(x) \Phi_{L}(x) + \cdots
\end{equation}
where the ellipsis denotes contributions suppressed by powers of $\lqcd/m_b$.
In this expression, the $B$ decay form factor $F^{\bar{B} \to H}$ and the light meson
decay constant $f_L$ are the nonperturbative parameters present in the case of
naive factorization (\ref{naive_factorization}).  The `non-factorizable' physics is
contained in the convolution of the perturbatively calculable hard-scattering kernel 
$T^I_i(x)$ with the light-cone momentum distribution of the leading
Fock state (quark-antiquark) of the light meson $\Phi_L(x)$.    
The parameter $x$ is the momentum fraction of one of the quarks inside the light meson.

To leading order in $\alpha_s$, one finds \cite{bbns2,politzer/wise91} 
\begin{equation}
\label{Tseries}
T^I_1(x) = 1 + {\cal O}(\alpha_s^2), \hspace{2cm} T^I_8(x) = 0 + {\cal O}(\alpha_s).
\end{equation}
Given that the light meson distribution function $\Phi_L(x)$ is normalized to unity,
naive factorization (\ref{naive_factorization}) is restored as the leading term
in a perturbative expansion. 

There are other decay topologies (penguin, annihilation) for which 
certain assumptions leading to (\ref{masterequation2}) 
are invalid.  However, detailed arguments show that these topologies are 
suppressed by powers of $\lqcd/m_b$, and are 
therefore irrelevant in the heavy quark limit \cite{bbns2}.

In the next section, we present the renormalon analysis of the `non-factorizable' 
corrections.  This will involve studying the large-order perturbative properties 
of (\ref{Tseries}) as a means of determining
the order at which power corrections enter the factorization equation 
(\ref{masterequation2}).

\section{The calculation}
\label{calculation}

For a particular heavy final state $H$ $(D,D^*)$ and light final state $L$ $(\pi,\rho)$ 
we must calculate the matrix elements
\begin{equation}
\langle {\cal O}_{1,8} \rangle \equiv \langle H(p') L(q) | {\cal O}_{1,8} | \bar{B}(p) \rangle.
\end{equation}

\begin{figure}[htbp]
\centerline{\scalebox{0.8}{\includegraphics{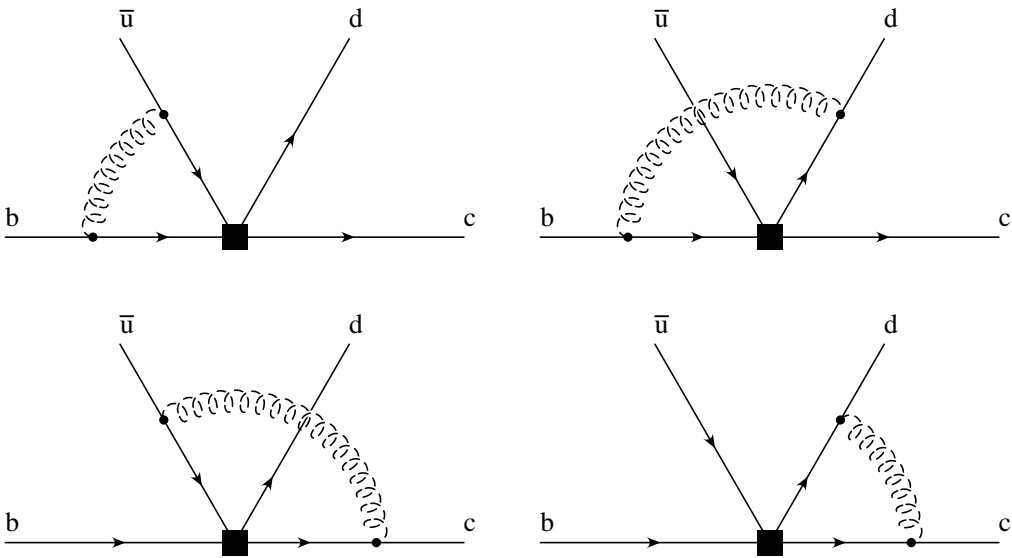}}}
\caption{The factorization-breaking corrections with renormalon propagators.} 
\label{diagrams} 
\end{figure} 

The matrix element for the singlet operator is the simplest, 
so we consider it first.  At leading order in $\alpha_s$
the operator factorizes cleanly into a product of currents.  The leading 
factorization-breaking QCD corrections are shown in Figure \ref{diagrams}. 
In order to create a color singlet structure from the resulting graph, 
one would have to consider the light meson to be in a Fock state higher than 
$(q' \bar{q})$; this situation, however, is power suppressed by $\lqcd/m_b$ \cite{bbns2}.
Alternatively, one could retain the leading Fock state and consider the exchange of 
two gluons rather than just one; this situation is suppressed by $\alpha_s$ relative
to the graphs in Figure \ref{diagrams}. 
In our calculation, then, the singlet matrix element is simply
\begin{eqnarray}
\langle {\cal O}_1 \rangle &=& \langle L(q) | \bar{d} \gamma^\mu (1 - \gamma_5) u | 0 \rangle \; \langle H(p') | 
	\bar{c} \gamma_\mu (1-\gamma_5) b | \bar{B}(p) \rangle \nonumber \\ 
 &=& i f_L \Big( \langle J_V \rangle - \langle J_A \rangle \Big)
\end{eqnarray}
where we define two matrix elements 
$\langle J_V \rangle = \langle H(p') | \bar{c} \gamma_\mu q^\mu b | \bar{B}(p) \rangle$ and 
$\langle J_A \rangle = \langle H(p') | \bar{c} \gamma_\mu q^\mu \gamma_5 b | \bar{B}(p) \rangle$.

The octet matrix element $\langle {\cal O}_8 \rangle$ has a richer structure.  It vanishes at leading order, but the
color structures of the graphs shown in Figure \ref{diagrams} are such that $\langle {\cal O}_8 \rangle$
receives perturbative corrections at all higher orders.  Let us denote the individual amplitudes of the diagrams in Figure \ref{diagrams} as ${\cal A}_i$, 
and the corresponding Borel transformed amplitudes as $B[{\cal A}_i]$.  
Defining $x$ to be the momentum fraction carried by the up quark of the light meson, $\bar{x} = 1-x$ 
to be the momentum fraction of the down quark, and $z = m_c/m_b$, the individual Borel transformed 
amplitudes of the diagrams shown in Figure \ref{diagrams} may be written as
\begin{eqnarray}
\label{bgraph1} B[{\cal A}_1](u) &=& \frac{i f_L C_F}{2 N_c} \left( \frac{\mu^2}{e^C m_b^2} \right)^u \left( F_1(x,z,u) \langle J_V \rangle - F_1(x,-z,u) \langle J_A \rangle \right) \nonumber \\
\label{bgraph2} B[{\cal A}_2](u) &=& \frac{i f_L C_F}{2 N_c} \left( \frac{\mu^2}{e^C m_b^2} \right)^u \left( F_2(\bar{x},z,u) \langle J_V \rangle - F_2(\bar{x},-z,u) \langle J_V \rangle \right) \nonumber\\
\label{bgraph3} B[{\cal A}_3](u) &=& \frac{i f_L C_F}{2 N_c} \left( \frac{\mu^2}{e^C m_c^2} \right)^u \left( F_2(x,1/z,u) \langle J_V \rangle - F_2(x,-1/z,u) \langle J_A \rangle \right) \nonumber \\
\label{bgraph4} B[{\cal A}_4](u) &=& \frac{i f_L C_F}{2 N_c} \left( \frac{\mu^2}{e^C m_c^2} \right)^u \left( F_1(\bar{x},1/z,u) \langle J_V \rangle - F_1(\bar{x},-1/z,u) \langle J_A \rangle \right)
\end{eqnarray}
where
\begin{eqnarray}
F_1(x,z,u) &=& \frac{\Gamma(1-2u)\Gamma(u-1)}{\Gamma(2-u)(x (1-z^2))^u} 
	\Bigg\{ \frac{(1-2u)(3u-1)}{u} f_1(x,z,u) \nonumber \\
&-& \left[ 1 - u - 2 \left( 1 - \frac{1}{x(1-z^2)} \right) \right] f_2(x,z,u) \ \Bigg\}
\end{eqnarray}
and
\begin{eqnarray}
F_2(x,z,u) &=& \frac{\Gamma(2-2u) \Gamma(u-1)}{\Gamma(3-u) (x(1-z^2))^u}
	\Bigg\{ \left[ \frac{2 - u + u^2 }{u} + \frac{2 u z}{1 - x(1-z^2)} \right]
	f_1(x,z,u) \nonumber \\
&+& \left[ \frac{2-u}{1-2u} \left( \frac{2}{x(1-z^2)} - (1+u) \right) - \frac{2 u z}{1 - x(1-z^2)}  \right] f_2(x,z,u) \Bigg\}.
\end{eqnarray}
We have written our result in terms of the hypergeometric functions
\begin{eqnarray}
f_1(x,z,u) &\equiv& _2F_1\left(1-u, u; 2-u;1 - \frac{1}{x(1-z^2)} \right) \\
f_2(x,z,u) &\equiv& _2F_1\left(1-u, 1+u; 2-u; 1 - \frac{1}{x(1-z^2)} \right).
\end{eqnarray} 

We have evaluated each of these graphs in $d = 4$ dimensions, as the divergences which would normally 
be present are regulated by the Borel parameter $u$ in the renormalon propagator (\ref{renProp}).  In the limit
$u \to 0$, however, each of the $B[{\cal A}_i](u)$ contain both infrared and ultraviolet divergences.  It was a 
central result of \cite{bbns1,bbns2} to show that in the heavy quark limit the sum
\begin{equation}
B_0[{\cal A}](u) = \sum_{i=1}^4 B[{\cal A}_i](u)
\end{equation}
is infrared finite for amplitudes of the type we are considering. This cancellation may be seen explicitly from our
amplitudes (\ref{bgraph4}):  in the vicinity $u \sim 0$ we have
\begin{eqnarray}
B[{\cal A}_1](u) &=& \frac{i f_L C_F}{2 N_c} \langle J_{V-A} \rangle \left[ \frac{1}{u^2} + 
\frac{1}{u}\left( -2 \log x (1-z^2) -C+ \log\frac{\mu^2}{m_b^2} - 2 \right) \right] + \cdots \nonumber \\
B[{\cal A}_2](u) &=& \frac{i f_L C_F}{2 N_c} \langle J_{V-A} \rangle \left[ -\frac{1}{u^2} + 
\frac{1}{u} \left( 2 \log \bar{x} (1-z^2) +C- \log\frac{\mu^2}{m_b^2} - 1 \right) \right] + \cdots \nonumber \\
B[{\cal A}_3](u) &=& \frac{i f_L C_F}{2 N_c} \langle J_{V-A} \rangle \left[ -\frac{1}{u^2} + 
\frac{1}{u} \left( 2 \log x (1-1/z^2) +C- \log\frac{\mu^2}{m_c^2} - 1 \right) \right] + \cdots \nonumber \\
B[{\cal A}_4](u) &=& \frac{i f_L C_F}{2 N_c} \langle J_{V-A} \rangle \left[ \frac{1}{u^2} + 
\frac{1}{u}\left( -2 \log \bar{x} (1-1/z^2) -C+ \log\frac{\mu^2}{m_c^2} - 2 \right) \right] + \cdots
\end{eqnarray}
where the ellipses denote terms finite as $u \to 0$, and we introduce the shorthand $\langle J_{V-A} \rangle = \langle J_{V} \rangle - \langle J_{A} \rangle$.  The collinear 
divergences cancel in pairs among the even and odd numbered amplitudes, while 
the remaining soft divergences cancel in the manner prescribed by Bjorken's 
color transparency argument \cite{bbns2,bjorken89}.

For the sum of the four amplitudes we find
\begin{equation}
\label{UVdivergence}
B_0[{\cal A}](u \sim 0) = -\frac{i f_L C_F}{2 N_c} \langle J_{V-A} \rangle 
	 \frac{6}{u} + \mbox{finite}.
\end{equation}
This remaining divergence is ultraviolet in origin, and may be renormalized in a 
manner consistent with MS-like subtraction schemes.
To this end, we follow the prescription of \cite{beneke/braun94,ball/beneke/braun95} by defining a renormalized amplitude
\begin{equation}
B[{\cal A}](u) = B_0[{\cal A}](u) + S_{\cal A}(u)
\end{equation}
where $S_{\cal A}(u)$ contains a divergence which cancels that in $B_0[{\cal A}](u)$ at the origin but is finite 
elsewhere. This regulating function may be written as
\begin{equation}
S_{\cal A}(u) = \frac{1}{u} \sum_{n=0}^\infty \frac{g_n}{n!} u^n
\end{equation}
where the coefficients $g_n$ are the expansion coefficients of another function $G(\epsilon) = \sum_{n=0} g_n \epsilon^n$
related to the amplitudes computed in $d = 4 - 2 \epsilon$ dimensions. We refer the reader to the relevant 
literature for an explanation of this method \cite{beneke/braun94,ball/beneke/braun95}. For the sum of amplitudes ${\cal A}$ we find
\begin{eqnarray}
G(\epsilon) &=& \frac{i f_L C_F}{2 N_c} \frac{ 2(1+\epsilon) (1 + 2 \epsilon) (3 + 2 \epsilon) \Gamma(4 + 2 \epsilon)}{ 3\, \Gamma(1 - \epsilon) \Gamma^2( 2 + \epsilon) \Gamma(3 + \epsilon) }  \langle J_{V-A} \rangle  \nonumber \\
&=& \frac{i f_L C_F}{2 N_c} \langle J_{V-A} \rangle (6 + 23 \; \epsilon + \frac{127}{6} \epsilon^2 + \cdot \cdot \cdot ).
\end{eqnarray}
The first term in this expansion cancels the ultraviolet divergence remaining in (\ref{UVdivergence}); $B[{\cal A}](u)$ is
infrared and ultraviolet finite. 

In this notation, the Borel transform of the factorization equation (\ref{masterequation2})
is
\begin{equation}
B[\langle {\cal O}_8 \rangle](u) = \int_0^1\,dx\,B[{\cal A}](u) \Phi_L(x).
\end{equation}
Comparing to (\ref{masterequation2}), we see that $B[{\cal A}](u)$ is proportional
to the Borel transform of the hard scattering kernel $T^I_8(x)$.

\subsection{Borel poles and power corrections}

Recall from (\ref{ambiguity}) that it is the pole nearest the origin on the positive real axis in the Borel plane which indicates the 
leading power correction.  A pole at the origin would indicate an ${\cal O}(1)$ correction, but the renormalization 
procedure outlined in the previous section ensures that no such pole is present.  We find that the first pole is located
at $u = 1/2$, corresponding to a power correction of ${\cal O}(\lqcd/m_b)$. This is not a very
surprising result, as power corrections of this order are known to be present from a variety of sources 
\cite{bbns2}. However, in the vicinity of this pole the Borel transformed amplitude 
has the form
\begin{equation}
\label{upole}
B[\langle {\cal O}_8 \rangle](u \sim 1/2) \propto \frac{1}{u-1/2} \int_0^1\,dx\,\Phi_L(x) \frac{(x-\bar{x})}{x \bar{x}} \langle J_{V-A} \rangle + \cdot \cdot \cdot
\end{equation}  
where $\bar{x} = 1-x$, and the ellipsis denotes nonsingular terms.

Before interpreting this result, we must specify the form of the light meson momentum
distribution $\Phi_L(x)$.  It is customary to write it as \cite{brodsky/lepage80}
\begin{equation}
\label{wfdecomposition}
\Phi_L(x) = 6 x (1-x) \left[ 1 + \sum_{n=1}^\infty \alpha_n^L(\mu) C_n^{3/2}(2x-1) \right]
\end{equation}       
where the Gegenbauer polynomials $C_n^{3/2}(y)$ are given by
\begin{equation}
C_n^{3/2}(y) = \left. \frac{1}{n!} \frac{d^n}{dh^n}(1 - 2 h y + h^2)^{-3/2}\right|_{h=0}.
\end{equation}

The distribution function $\Phi_L(x)$ is a nonperturbative
object for which the Gegenbauer moments $\alpha_n^L(\mu)$ are unspecified.  
It is known, however, that 
$\alpha_n^L(\mu \to \infty) = 0$ \cite{brodsky/lepage80}. For $\mu \sim m_b \gg \lqcd$,  
one may take the distribution
function to have the asymptotic form $\Phi_L^0(x) = 6 \, x (1-x)$ up to power corrections
of ${\cal O}(\lqcd/m_b)$.  
In this case the wavefunction is symmetric under $x \to 1-x$,
and the integration over $x$ in (\ref{upole}) vanishes, removing the pole.  

More generally, note that 
the Gegenbauer polynomials $C_{n}^{3/2}(2x-1)$ in (\ref{wfdecomposition}) 
with $n$ even are, like $\Phi_L^0(x)$,  even under $x \to 1-x$, 
while those
with $n$ odd are odd under the same replacement.  The result of the integration (\ref{upole}) may
then be written as
\begin{equation}
\label{upoleOdd}
B[\langle {\cal O}_8 \rangle](u \sim 1/2) \propto \frac{6}{u-1/2} \langle J_{V-A} \rangle 
	\left( \alpha_1^L(\mu) + \alpha_3^L(\mu) + \alpha_5^L(\mu) + \cdots \right).
\end{equation}

 Only the asymmetric terms of $\Phi_L(x)$ contribute to the residue. Thus
the renormalon analysis indicates the presence of ${\cal O}(\lqcd/m_b)$
factorization-breaking power corrections only for asymmetric light meson
wavefunctions.
 
For certain final states of interest $(L = \pi,\rho)$, SU(2) symmetry
ensures that the wavefunction is symmetric
\cite{ball/braun96,bagan/ball/braun97}. In these cases the pole at $u =
1/2$ disappears, and the renormalon analysis gives no indication of
factorization-breaking ${\cal O}(\lqcd/m_b)$ corrections, suggesting that
such power corrections are likely absent.  This result implies that power
corrections due to the `non-factorizable' vertex diagrams in Figure
\ref{diagrams} should be suppressed in $\bar{B} \to D^{(*)+} \pi^-$ and
$\bar{B} \to D^{(*)+} \rho^-$ decays relative to decays into a light meson
with an asymmetric distribution function, such as $\bar{B} \to D^{(*)+}
K^-$. 

In assessing this result, one should keep in mind that there are 
other sources of power corrections to these decays which
our analysis does not address.  For example, we have not considered
soft interactions with the spectator quark, annihilation diagrams,
or interactions with sub-leading Fock states involving
additional soft partons \cite{bbns1,bbns2}. Our conclusions about the vanishing 
${\cal O}(\lqcd/m_b)$ corrections in the symmetric limit apply only
to the class of diagrams we have considered.
 
The next renormalon pole is located at $u=1$, corresponding to an
${\cal O}(\lqcd^2/m_b^2)$ correction, and here there is a nonzero residue
even in the case of a symmetric wavefunction $\Phi_L(x)$. Therefore one
should expect power corrections of this order for all decay modes. 

\section{Conclusions}
\label{conclusions}

 In this paper we have carried out a renormalon analysis of factorization-breaking 
effects in $\bar{B}$ meson decays to certain hadronic heavy-light final states.
The renormalon approach, which probes the theory at high orders in perturbation 
theory, allows us to learn about nonperturbative power corrections.
In the low energy effective theory governing the $\bar{B}$ decays there are 
two operators, a color singlet and a color octet. 
The factorization-breaking corrections to the singlet operator are, however, suppressed by
powers of $\alpha_s$ or $\lqcd/m_b$ relative to the octet operator; for this reason we
focus our analysis on the octet.  

We find that the renormalon analysis of `non-factorizable' corrections 
to the octet matrix element indicates the presence of power 
corrections of ${\cal O}(\lqcd/m_b)$. 
We also find that these leading power corrections are sensitive 
only to asymmetries in the light meson light-cone parton distribution function. 
Thus for a symmetric distribution function the leading renormalon
pole vanishes, suggesting the absence of ${\cal O}(\lqcd/m_b)$ `non-factorizable' 
corrections in such cases.   
The next power corrections are present at ${\cal O}(\lqcd^2/m_b^2)$.  

As it is natural to expect the light meson wave function to be symmetric for 
certain final states $(L = \pi,\rho)$, 
the potential vanishing of the ${\cal O}(\lqcd/m_b)$ power corrections in the 
symmetric limit is an 
interesting result that warrants a few additional comments.  First, it should be 
noted that a particular decay $\bar{B} \to H L$ of the type we have
considered in this paper always involves a combination of the singlet and 
octet operator matrix elements.  Our result, however, applies only to the 
octet matrix element --- we are unable to conclude from our analysis whether
the leading power corrections to the singlet matrix element vanish in an 
analogous way.

Second, the power corrections we can probe through the renormalon analysis
are physically due to soft and collinear `non-factorizable' gluon exchange, 
and our comments about suppression of
the leading corrections should be understood as referring to these effects only.  
There are, however, other ${\cal O}(\lqcd/m_b)$ corrections originating 
from diagrams 
which were not considered in our analysis.  
A more detailed examination of 
these effects would have to be undertaken to accurately assess the size 
of ${\cal O}(\lqcd/m_b)$ corrections to a physical decay amplitude.    

\acknowledgments
We would like to thank Michael Luke for many discussions related to
this project, and Christian Bauer and Pierre Savaria for their comments on the manuscript. 
This work is supported by the Natural Sciences and Engineering Research Council of Canada, 
and by the Walter B. Sumner Foundation.

\end{document}